# Introduction of spin centers in single crystals of $Ba_2CaWO_{6-\delta}$

Mekhola Sinha,[1] Tyler J. Pearson,[2] Tim R. Reeder,[3] Hector K. Vivanco,[1] Danna E. Freedman,[2] W. Adam Phelan,[1] and Tyrel M. McQueen* [1,3,4]

[1] Department of Chemistry, The Johns Hopkins University, Baltimore, MD 21218

[2] Department of Chemistry, Northwestern University, Evanston, IL 60208

[3] Department of Physics and Astronomy and the Institute for Quantum Matter, The Johns Hopkins University, Baltimore, MD 21218

[4] Department of Materials Science and Engineering, The Johns Hopkins University, Baltimore, MD 21218

**ABSTRACT:** Developing the field of quantum information science (QIS) hinges upon designing viable qubits, the smallest unit in quantum computing. One approach to creating qubits is introducing paramagnetic defects into semiconductors or insulators. This class of qubits has seen success in the form of nitrogen-vacancy centers in diamond, divacancy defects in SiC, and P doped into Si. These materials feature paramagnetic defects in a low nuclear spin environment to reduce the impact of nuclear spin on electronic spin coherence. In this work, we report single crystal growth of $Ba_2CaWO_{6-\delta}$, and the coherence properties of introduced $W^{5+}$ spin centers generated by oxygen vacancies. $Ba_2CaWO_{6-\delta}$ ($\delta$ = 0) is a B-site ordered double perovskite with a temperature-dependent octahedral tilting wherein oxygen vacancies generate $W^{5+}$ ($d^1$), $S$ = ½, $I$ = 0, centers. We characterized these defects by measuring the spin-lattice ($T_1$) and spin-spin relaxation ($T_2$) times from T = 5 to 150 K. At T = 5 K, $T_1$ = 310 ms and $T_2$ = 4 μs, establishing the viability of these qubit candidates. With increasing temperature, $T_2$ remains constant up to T = 60 K and then decreases to $T_2$ ~ 1 μs at T = 90 K, and remains roughly constant until T = 150 K, demonstrating the remarkable stability of $T_2$ with increasing temperature. Together, these results demonstrate that systematic defect generation in double perovskite structures can generate viable paramagnetic point centers for quantum applications and expand the field of potential materials for QIS.

## I. INTRODUCTION

The advent of quantum information science (QIS) will fundamentally change our approach to computation, allowing us to answer currently intractable questions in a myriad of fields including biochemistry, physics, and cryptography [1,2]. QIS is an aggregate term comprising quantum sensing, quantum computing, and quantum metrology amongst other fields. Of these areas, the most advanced is quantum computing which relies on a computing paradigm wherein information is processed using quantum bits, or qubits, which can be placed into an arbitrary superposition of two states. A wide range of systems from photons to superconducting devices have been studied as candidates for QIS [3].

A highly promising area of research is harnessing electronic spins as qubits. As inherently quantum systems, electronic spins are ideal qubit candidates both for their modularity and their ease of manipulation with microwave radiation [4,5,6,7,8,9,10,11]. Electronic defect sites, in particular nitrogen-vacancy sites in diamond [12], phosphorus defects in silicon [5], and double-vacancy sites in silicon carbide are prominent examples [13,14,15]. However, to create systems with the longest possible coherence times, we must continue to glean new insights into what drives decoherence and develop new materials design principles for qubit hosts. $T_1$, the longitudinal electronic spin relaxation time constant, relates to the spin-lattice relaxation time of the electronic spin. This parameter represents the maximum data storage time of an electronic spin. $T_2$ is the spin echo dephasing time constant in the *xy*-plane and relates to the spin-spin relaxation time. $T_2$ also represents the coherence time, the maximum operation time of a qubit [16]. The factors which influence these parameters are different, so to maximize the coherence time in a system, one must carefully design around both $T_1$ and $T_2$. Recent studies of $T_2$ both in molecular and in solid-state systems have found that nuclear spins play a significant role in promoting decoherence [7,9,10,11,15,17,18]. As such, removal of nuclear spin has become one of the most widely adopted design parameters for electronic spin qubit systems [6,19,20,21,22]. While fundamentally, $T_2$ represents the functional operating time of a qubit, we often find that $T_1$ is the most restrictive parameter in practice, as $T_1$ represents the theoretical upper limit to $T_2$. The chemical properties leading to maximization of $T_1$ remain an open question and indeed many recent studies have focused heavily on questions related to $T_1$ [23,24,25]. Recent advances in pulse decoupling techniques have demonstrated that a short $T_2$ can be overcome [26,27,28,29], but $T_1$ remains a limiting factor. Since the longitudinal relaxation time, $T_1$, represents the upper limit to the phase memory time, we must find ways to control the processes leading to $T_1$ relaxation.

In this manuscript, we test a design strategy to avoid electronic excitations and destruction of spin information by utilizing a wide band gap insulating oxide as a host material. In the case of a wide band gap material, the difference between the ground and excited states is large enough to avoid band-to-band and defect-to-band transitions, thereby causing native defects to produce trap states known as deep traps far away from the band edges. This allows certain defects, such as nitrogen vacancy centers, to be initialized and measured at room temperature [30]. Further, for the materials system used, it is possible to systematically introduce single-spin isolated paramagnetic centers by controlled defect chemistry.

We report the single crystal synthesis and structure of $Ba_2CaWO_{6-\delta}$, an insulating double perovskite with a large band gap (3.6 eV) [31]. The primary naturally occurring isotopes of Ba, Ca, W and O have zero nuclear spin, largely eliminating spin-bath effects from the lattice. Furthermore, the oxide framework itself is rigid, helping suppress local vibrationally mediated decoherence mechanisms. The introduction of oxygen vacancies creates nominally $W^{5+}$ ($d^1$), $S = \frac{1}{2}$, $I = 0$ point defects in the single crystals of $Ba_2CaWO_{6-\delta}$. Coherence studies of $Ba_2CaWO_{6-\delta}$ show promising longitudinal relaxation time $T_1$ and relatively constant transverse relaxation time $T_2$ over a temperature range of T = 20 to 60 K, lending credence to our design strategy.

## II. EXPERIMENTAL SECTION

### A. Single crystal growth

$Ba_2CaWO_6$ powder was purchased from Sigma Aldrich (99.9%). The purchased powder was sealed in a rubber tube, evacuated, and compacted into a rod (typically 5 mm in diameter and 60 mm long for the feed and 25 mm long for the seed) using a hydraulic press under an isostatic pressure of 70 MPa. After removal from the rubber tube, the rods were sintered at T = 1000 °C for 24 hours in air. A Laser Diode Floating Zone (LDFZ) furnace (Crystal Systems Inc FD-FZ-5-200-VPO-PC) with 5 × 200 W GaAs lasers (976 nm) was used as the heating source. During all of the growths, the molten zone was moved upwards with the seed crystal being at the bottom and the feed rod above it. This was accomplished by holding the lasers in a fixed position at an angle of 4° to the horizontal axis and translating both the seed and feed rods downwards. Successive optimizations yielded final parameters of 15 mm/hr feeding rate, 10 mm/hr growing rate and 20 rpm rotation rate for counter-rotating rods.

### B. Characterization

Synchrotron X-ray powder diffraction data (XRPD) were obtained from 11-BM-B at Argonne National Laboratory using a Bending Magnet (BM) of critical energy 19.5 keV as the source. Twelve independent analyzers each separated by ~ 2° in 2θ and consisting of a Si (111) crystal and a $LaCl_3$ scintillator where used for detection. Backscattered X-ray Laue diffraction (with the X-ray beam of about 1 mm in diameter) was utilized to check the orientations of the crystals. The microstructural homogeneity of sample surfaces cut directly from the cross sections of the as-grown crystals was probed using a JEOL JSM IT100 scanning electron microscope (SEM) at 20 keV operating in backscatter mode. Laboratory-based X-ray diffraction patterns were collected using a Bruker D8 Focus diffractometer with $CuK_\alpha$ radiation and a Bruker D8 Advance with an Oxford Cryosystem PheniX cryocontroller with $CuK_\alpha$ radiation from T = 80 – 300 K. Phase identification and unit cell determinations were carried out using the Bruker TOPAS software (Bruker AXS). A Quantum Design Physical Properties Measurement System (PPMS) was used for the heat capacity measurements from T = 1.9 to 300 K at $\mu_o H$ = 0 T using the semi-adiabatic method. Electron paramagnetic resonance (EPR) spectroscopy was performed on crushed microcrystalline powders contained within a 4 mm OD quartz EPR tube. EPR data were obtained at T = 297 K at X-band frequency (~0.3 T, 9.5 GHz) on a Bruker E580 X-band spectrometer equipped with a 1 kW TWT amplifier (Applied Systems Engineering) and on a Bruker E580 X-band spectrometer at the National Biomedical EPR Center at the Medical College of Wisconsin (Milwaukee, WI) equipped with a 1 kW TWT amplifier. Temperature was controlled using an Oxford Instruments CF935 helium cryostat and an Oxford Instruments ITC503 temperature controller (UIUC), and an Oxford Instruments ITC503S temperature controller (MCW). All data were processed using a combination of Xepr, Python 2.7, Origin Pro 2015, and MatLab R2018b. EasySpin [32] was used to simulate the CW (continuous wave) EPR spectra of the W 5$d$ ions.

## III. Results and Discussion

### A. Single crystal growth

$Ba_2CaWO_6$ is a B-site ordered double perovskite with a twelve coordinate Ba cation site and octahedrally coordinated smaller cation Ca/W sites consisting of alternate large and small octahedra of Ca and W, respectively, see Fig 1(a). The high vapor pressures of BaO and CaO at the melting temperature of $Ba_2CaWO_6$ (T ≈ 1450 °C as determined from thermogravimetric analysis) makes it difficult to grow crystals via direct melting at atmospheric pressure [33]. Initial floating zone crystal growth attempts in oxygen, air, carbon dioxide, and static argon atmospheres at atmospheric pressure led to significant vaporization of calcium (in the form of oxides), and, as a result, produced samples containing a $BaWO_4$ impurity phase, leading to cracks. The optimum growth conditions were found to be a 7 bar high-purity argon atmosphere flowing at the rate of 2.5 mL/min and 40% of laser power. These conditions significantly reduce the vaporization of calcium oxide, which allowed us to obtain blue tinted single crystals. A typical $Ba_2CaWO_{6-\delta}$ crystal (45 mm in length and 5 mm in diameter) grown using optimal conditions is shown in Fig 1(a). X-ray Laue photographs along the cross section of the grown crystals show no detectable variation in orientation and no evidence of twinning, see Fig 1(b).

A back-scattered SEM micrograph taken from a cross section of a crystal complements our Laue data showing a uniform microstructure with no evidence of domain formation or any inclusions on the micrometer length scale, see Fig 1(c). The as-grown crystals are blue (a signature of $W^{5+}$) and oxygen deficient, a result of the reducing atmosphere. The precise oxygen vacancy concentration is controllable via post-annealing; single crystals were sintered in flowing oxygen for one month at T = 1050 °C to obtain colorless single crystals of $Ba_2CaWO_6$, demonstrating tunability of the oxygen vacancy concentration, and hence the number of spin centers, see Fig 1(d).

### B. Crystal structure

To understand the origin and behavior of the relaxation rates, we analyzed the structure of the grown crystals and the thermal properties of the material in more detail. The structures of as grown $Ba_2CaWO_{6-\delta}$ single crystal were solved in space group *I2/m* at T = 295 K and at 100 K via Rietveld refinement to synchrotron XRPD. We find that introduction of Ba site disorder and replacement of about 14% Ba atoms with Ca into the structural model result in significant improvement of the Rietveld fit. No temperature dependent structural phase transition was observed within measurement resolution. The structural model obtained shows that at T = 295 K the structure consists of an octahedral tilt pattern ($a^-b^-c^0$) with a W-O1-Ca angle of 173° (a- axis tilt) and a W-O2-Ca angle of 167° (b- axis tilt) [34]. At T = 100 K W-O1-Ca angle becomes 163° and W-O2-Ca angle is 169°. The deviations from pseudo-cubic structure (distortion index i.e. standard deviations of the monoclinic lattice parameters from the pseudo-cubic lattice parameters) were calculated using cell parameters obtained from Le Bail fits to laboratory XRPD data collected at different temperatures with high purity Si as standard. Fig 2 shows that the average structure is closest to pseudo-cubic at T = 230 K and deviates further from cubic with increasing or decreasing temperature due to in-plane rotations of the Ca/WO$_6$ octahedra. Such structural flexibility is expected to give rise to low-lying optical phonon modes that may increase spin-lattice relaxation *via* phonon-mediated processes.

Specific heat capacity measurements support this hypothesis. Figure 3 shows the heat capacity for a single crystal of $Ba_2CaWO_{6-\delta}$ as $C_p/T^3$ vs log T. These data approximate the one-dimensional phonon density of states. Modeling the data illustrates the various contributions to the phonon density of states [35]. In the model we used, a Debye (acoustic) contribution is represented by a constant up to $\theta_D$, whereas an Einstein (optical) mode is represented by a peak, resulting from activated low T behavior. The most striking feature of the plot is a large low temperature peak at T = 21 K with $\theta_E$ = 105 K, indicating the data cannot be described by any combination of

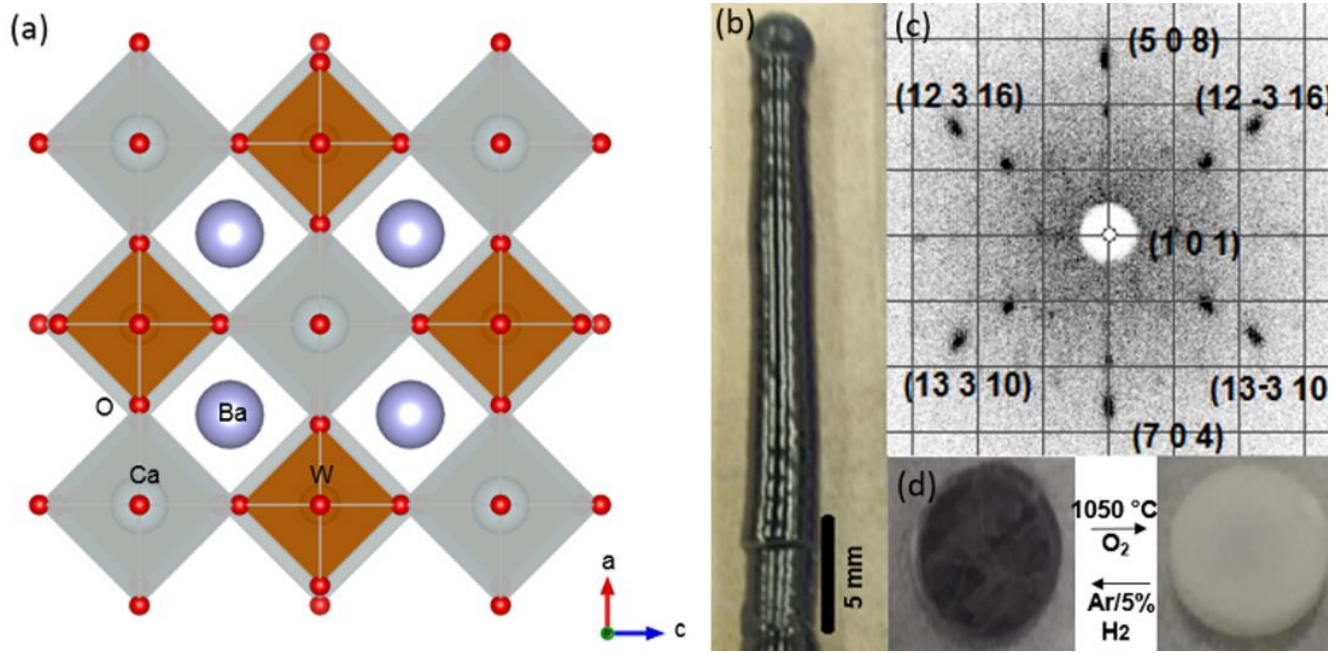

**Figure 1.** (a) Double perovskite structure of $Ba_2CaWO_6$ (δ = 0) containing alternate octahedra of WO$_6$ (orange) and CaO$_6$ (grey) and twelve coordinate Ba atom in the A site. The isolated WO$_6$ octahedra are primed to become spin centers upon introduction of electrons by removal of oxygen (red spheres). (b) A typical $Ba_2CaWO_6$ single crystal. (c) Representative Laue diffraction along the (101) direction, perpendicular to the growth axis. (d) Oxygen vacancy concentration can be tuned by sintering under controlled oxygen partial pressure.

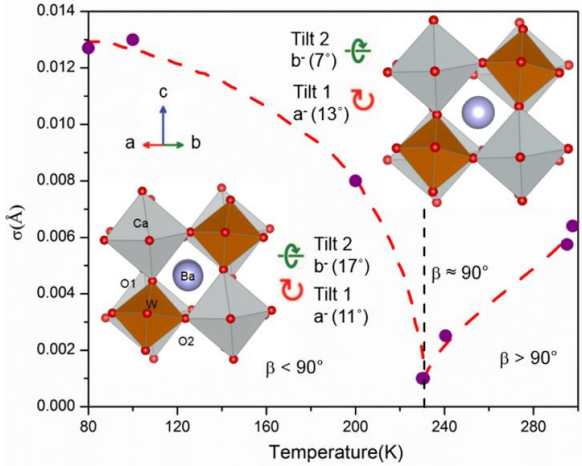

**Figure 2.** Standard deviations from the pseudo-cubic lattice parameters (distortion index) for single crystal of $Ba_2CaWO_{6-\delta}$ with increasing temperature where β is the angle between a and c axes. A line to guide the eye (red) shows the distortion index starts decreasing around T = 140 K, it reaches closest to the pseudo-cubic structure at T = 230 K (black line) before going up again. Crystal structure of $Ba_2CaWO_{6-\delta}$ obtained from refined synchrotron XRPD data collected at T = 295 K and T = 100 K shows the double perovskite structure containing alternate octahedra of $CaO_6$ (grey) and $WO_6$ (orange). The structure shows an octahedral tilt pattern of (a⁻b⁻c⁰) at each temperature. This octahedral distortion is expected to activate low lying phonon modes in the structure resulting in faster relaxation of the spin.

Debye terms. We observed that this low-lying Einstein phonon mode becomes operative above T = 10 K, resulting in a higher phonon density of states above T = 15 K. This corresponds to the onset of the local process as determined by modeling the temperature dependence of $T_1$. Low energy optical modes can usually be attributed to a specific structural feature. It is likely the low energy vibrational states are associated with large amplitude motion of the oxygen atoms in the $Ca/WO_6$ octahedra resulting in distortion of the structure from the pseudo-cubic lattice. Two-thirds (0.67) of the entropy of the Einstein mode is recovered by T = 230 K where the soft phonon modes are excited such that the average becomes pseudo-cubic. While the low-lying Einstein peak can be described by the distortion in the local modes, the high temperature Debye mode is associated with higher energy vibrations of atoms having lower masses. Inversely, the lower temperature Debye mode is associated with lower energy vibrations of atoms having higher atomic masses. The high energy part of the spectrum was fitted with two Debye contributions having $\theta_{D1}$ = 263 K and $\theta_{D2}$ = 793 K, a reasonable value for high energy modes in oxides. The total number of oscillators sums to 10.23(6) (the number of atoms per formula unit) which compares well to the theoretical value of 10. The values of the oscillator strengths and characteristic temperatures are given in Table I. The small upturn at T ≈ 2 K is due to helium condensation around this temperature.

**Table I:** Characteristic temperatures and number of oscillators used in fitting Einstein and Debye modes to the specific heat for single crystal of $Ba_2CaWO_{6-\delta}$.

| Mode | Temperature (K) | Oscillator strength /formula unit |
|---|---|---|
| Einstein | $\theta_E$ = 104.97(26) | 1.23(1) |
| Debye 1 | $\theta_{D1}$ = 263(3) | 3.5(1) |
| Debye 2 | $\theta_{D2}$ = 793(63) | 5.5(6) |

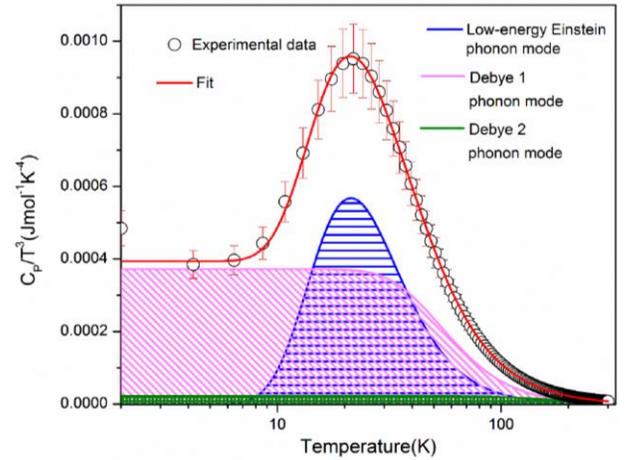

**Figure 3**. Heat capacity ($C_p$) divided by temperature cubed ($T^3$) versus log of temperature (T) for a single crystal piece of $Ba_2CaWO_{6-\delta}$. The red line shows a fit to the experimental data including the Einstein and Debye terms. Contributions of individual components are plotted below: pink – Debye 1 phonon mode heat capacity, green – Debye 2 phonon mode heat capacity, blue – low energy Einstein phonon mode. The low-lying Einstein peak describes the distortion in the local modes while the high temperature Debye mode is associated with higher energy vibrations due to atoms having lower masses and the lower temperature Debye mode is associated with lower energy vibrations of atoms having higher atomic masses.

### C. Determination of the spin-state of the defect center

To characterize the spin state of the magnetic W ions resulting from oxygen deficiencies ($Ba_2CaWO_{6-\delta}$), we utilized continuous wave and pulse EPR techniques. To determine the spin state of the system, we performed transient nutation experiments calibrated against a 2,2-diphenyl-1-picrylhydrazyl (dpph) radical standard. These allowed us to confirm that the spin state which we observed was a doublet and that the transition we observed was between the $M_S = \pm 1/2$ sublevels. With this information in hand, we could easily simulate the experimental cw spectrum using EasySpin [32] to a spin doublet (Figure 4) with $g$ = 1.96, which is in line with similar observations in oxygen-deficient tungsten oxides [36]. Taken together, this information leads us to conclude that the spin centers introduced by defect generation are $W^{5+}$ ions. The weakness of this transition likely arises from low spin concentration across the sample. The spectrum in this case was measured at T = 297 K. We attempted to utilize a model developed according to theory outlined by *Salikhov et al.* [37] to attempt to approximate spin concentration. However, the model cannot account for spin-spin interactions of more than approximately 50 Å. Therefore, we cannot absolutely say the concentration of spins in the sample, only that they are more than 50 Å apart (i.e. < 0.02%).

### C. Pulse EPR measurements

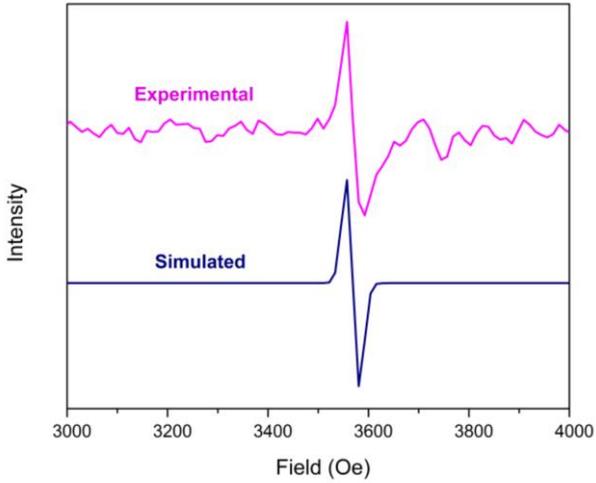

**Figure 4.** Comparison between the resonance peak obtained from experimental continuous wave spectra and the simulated fit to the experimental data to a spin doublet with *g* = 1.96 using EasySpin.

To investigate the feasibility of the grown crystals as qubit candidates and to learn more about how the electrons interact with the material, we performed pulse EPR spectroscopy on finely ground single crystals of $Ba_2CaWO_{6-\delta}$. The immediate aim of these measurements was to determine the parameters most relevant to understanding the coherence properties of the electronic spin: $T_1$ and $T_2$.

We determined $T_1$ across the temperature range T = 5 K to 150 K. We accomplished this using an inversion recovery pulse sequence ($\pi - T - \pi/2 - \tau - \pi - \tau -$ echo). We fit the resulting inversion recovery curves using an exponential function of the form

$$I(\tau) = -A \left( e^{-(\tau/T_1 + \sqrt{\tau/c})} - b - 1 \right)$$

where *I* is the normalized echo intensity, *A* and *b* are normalization factors (1 and 0 approximately), and *c* is a factor related to spectral diffusion, which is relevant at low temperatures in this system (Fig 5a). We found that at T = 5 K, $T_1$ is 310(30) ms in this material, which provides a high ceiling to the coherence time of the system. The $T_1$ of $W^{5+}$ centers in $Ba_2CaWO_{6-\delta}$ compares very well to some of the best qubit candidates known.

To investigate the spin-lattice relaxation in more detail, we examined and modeled the temperature dependence of $T_1$, see Fig 5(b). Modeling this dependence is useful for determining which processes mediate spin-lattice relaxation as each process has a unique and well-defined temperature dependence. The model we utilized to fit the data was

$$\frac{1}{T_1} = A_{dir}T + A_{ram1}\left(\frac{T}{\theta_{D1}}\right)^9 J_8\left(\frac{\theta_{D1}}{T}\right) + A_{ram2}\left(\frac{T}{\theta_{D2}}\right)^9 J_8\left(\frac{\theta_{D2}}{T}\right) + A_{loc}\frac{e^{\theta_E/T}}{(e^{\theta_E/T}-1)^2}$$

where $A_{dir}$, $A_{ram1}$, $A_{ram2}$ and $A_{loc}$ are coefficients reflecting the influence of the direct, Raman and the local processes respectively. Each Raman process is associated with a characteristic Debye temperature ($\theta_{D1}$ and $\theta_{D2}$), the local mode an Einstein temperature ($\theta_E$), fixed to values obtained from quantitative analysis of the heat capacity, and $J_8$ is the transport integral describing the joint phonon density of states assumed by the Debye model. This integral takes the form

$$J_8\left(\frac{\theta_D}{T}\right) = \int_0^{\theta_D/T} x^8 \frac{e^x}{(e^x - 1)^2} dx$$

This overall model for $T_1$ includes terms for the direct relaxation process, two Raman relaxation processes, and a term describing additional relaxation due to local modes. The direct process proceeds via a spin flip mechanism and is typically relevant only at approximately T = 10 K or below. The Raman and local processes are two-phonon processes analogous to Raman photon scattering and become operative at temperatures where there is substantial thermal phonon population. Examination of the temperature dependence of $T_1$ hints at why $T_1$ is long at T = 5 K in this system: the influence of the direct process is very small in this material, with $A_{dir}$ = 0.49(09) $K^{-1}s^{-1}$. Typical values for $A_{dir}$ fall within the approximate range of 10-50 $K^{-1}s^{-1}$[38]. With minimal thermal population of the phonon modes, the only means by which the spin can relax is through the direct process. However, at X-band (~0.3 T) and in the absence of a strong hyperfine field – a result of the mostly nuclear spin-free nature of this material in which the spin resides – the direct process is slow [35]. However, with the onset of the two-phonon processes between T = 15-20 K, the spin-lattice relaxation time begins to precipitously decrease. Generally, this decrease most likely arises from the presence of heavy elements with high spin-orbit coupling in the lattice. With increasing spin-orbit coupling, the strength of spin-phonon coupling also increases, leading to enhanced relaxation rates [38].

We can gain quantitative information about the relative strength of spin-phonon coupling to the different phonon contributions by utilizing knowledge of the oscillator strength from heat capacity combined with the coefficients from the fits to $1/T_1$. The coefficient for a two-phonon process will have the general form [39]:

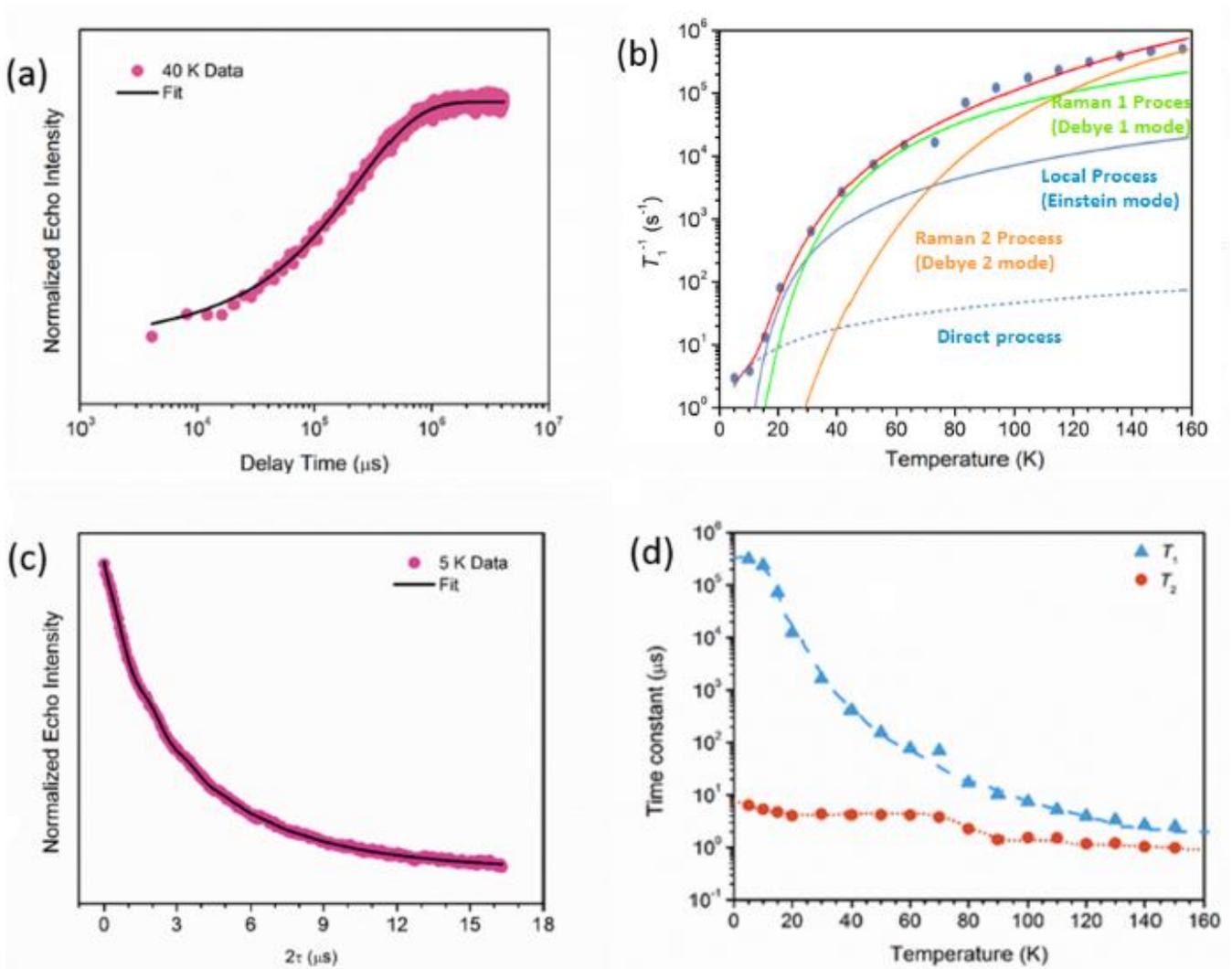

**Figure 5**. Pulse electron paramagnetic resonance spectroscopic data. (a) Inversion recovery curve used for calculating the spin-lattice relaxation time, $T_1$ at T = 40 K. (b) Temperature dependence plot of the rate of longitudinal relaxation $T_1^{-1}$ with a quantitative fit to a combination of a Direct, two "Raman" (Debye-phonon), and one "local" (Einstein-phonon) contributions. Parameters obtained from the fit (red line) are $A_{dir}$ = 0.49(09) K$^{-1}$s$^{-1}$, $A_{ram1}$ = 4.4(8)·10$^6$ s$^{-1}$, $A_{ram2}$ = 3.2(8)·10$^8$ s$^{-1}$, and $A_{loc}$ = 8(2)·10$^3$ s$^{-1}$. The associated Debye and Einstein temperatures were fixed at the values obtained from heat capacity measurements. (c) Hahn-echo decay curve used to calculate spin-spin relaxation time $T_2$. (d) Temperature dependence plot of $T_2$ and $T_1$. (Lines to guide the eye shows the trend followed by $T_1$ and $T_2$ with temperature).

$$\frac{1}{T_1} = \frac{4\pi\hbar^2}{\rho_{cryst}^2 v_S^4} \frac{\langle b|V^{(1)}|a\rangle^4}{\Delta_{cryst}^4} \int_0^{\omega_{max}} \bar{n}(\bar{n}+1)\,\omega^4\,\rho^2(\omega)\,d\omega$$

where $v_s$ is the average sound velocity, $\rho_{cryst}$ is the crystal density and $\Delta_{cryst}$ is the orbital splitting. The $\langle b|V^{(1)}|a\rangle \equiv G$ is the matrix element called spin-phonon coupling parameter where $V^{(1)}$ is the crystal field potential produced by phonons generating strain $\varepsilon$ and taken from the expansion of the potential into power series V = V$_o$ + $\varepsilon$ V$^{(1)}$ + $\varepsilon^2$ V$^{(2)}$ + ….. For Debye (Raman)-type phonon process, $\rho(\omega) = \frac{3 N_D \omega^2}{2 \pi^2 v_s^3}$, where N$_D$ is the number of oscillators related to the Debye mode, $v_s = \frac{k_B \theta_D}{\hbar \left(6\pi^2 \frac{N}{V}\right)^{\frac{1}{3}}}$

where N is the number of atoms in a crystal of volume V, resulting in:

$$\frac{1}{T_1} = A_{ram}\left(\frac{T}{\theta_D}\right)^9 J_8\left(\frac{\theta_D}{T}\right)$$

where $A_{ram} = \dfrac{9 \cdot 6^{\frac{10}{3}} \pi^{\frac{11}{3}} \hbar^3 \, N_D^2 \left(\frac{N}{V}\right)^{\frac{10}{3}} G_{ram}^4}{\rho_{cryst}^2 \, k_B \, \theta_D \, \Delta_{cryst}^4}$.

Similarly, for an Einstein (Local)-type phonon process, $\rho(\omega) = 3 \cdot N_E \left(\frac{N}{V}\right) \delta(\omega - \omega_e)$, $\nu_s = \frac{k_B \theta_E}{\hbar \left(\pi^3 \frac{N}{V}\right)^{\frac{1}{3}}}$, where $N_E$ is the number of oscillators related to the Einstein mode, resulting in:

$$\frac{1}{T_1} = A_{loc} \frac{e^{\theta_E/T}}{(e^{\theta_E/T} - 1)^2}$$

where $A_{loc} = \frac{36 \pi^5 \hbar^3 N_E^2 \left(\frac{N}{V}\right)^{\frac{10}{3}} G_{loc}^4}{\dot{\rho}_{cryst}^2 k_B \theta_E \Delta_{cryst}^4}$. Using these, the ratio of coefficients for two Raman-type processes is:

$$\frac{A_{ram2}}{A_{ram1}} = \left(\frac{N_{D2}}{N_{D1}}\right)^2 \left(\frac{\theta_{D1}}{\theta_{D2}}\right) \left(\frac{G_{ram2}}{G_{ram1}}\right)^4$$

Whereas the ratio of coefficient for a Raman and Local process is:

$$\frac{A_{loc}}{A_{ram}} = \frac{36 \pi^5}{9 \cdot 6^{\frac{10}{3}} \pi^{\frac{11}{3}}} \left(\frac{N_E}{N_D}\right)^2 \left(\frac{\theta_D}{\theta_E}\right) \left(\frac{G_{loc}}{G_{ram}}\right)^4$$

We find that the spin-phonon coupling to the Debye mode with $\theta_{D2}$ = 793(63) K (Debye 2) is 3.2(4) times greater than to the one with $\theta_{D1}$ = 263(3) (Debye 1). Similarly, the coupling to the Einstein mode is 1.1(7) relative to Debye 1. These provide evidence that the spin-phonon coupling is stronger for higher frequency oscillators in this material.

Next, we examined $T_2$ over the same temperature range (Table SIII). We utilized a two-pulse Hahn-echo sequence to monitor the stability of the superposition as a function of the inter-pulse delay time. We then fit the decay curves with a stretched exponential function of the form
$$I(\tau) = A\left(1 - B\cos(\omega\tau + d)e^{-\tau/T_{osc}}\right)e^{-(2\tau/T_2)^q} + f$$
see Fig 3(c). The function includes two parts, the first of which describes the electron spin-echo envelope modulation (ESEEM) within the decay curve and the second of which describes the spin-echo decay itself. Within the function as written, $A$ is a normalization factor (approximately 1), $B$ is the ESEEM modulation amplitude, $\omega$ is the ESEEM frequency in MHz, $d$ is the modulation phase, $T_{osc}$ is the ESEEM decay time, $q$ is the stretch factor, and $f$ is an offset term to assist in fitting (approximately 0). The inclusion of the ESEEM term was necessary to model the small oscillations within the decay curve resulting from the interaction between the electronic spin and the 14.3% abundant $I = {}^1/_2$ $^{183}$W nucleus. However, the data above T = 70 K was fitted to the equation
$$I(\tau) = Ae^{-(2\tau/T_2)^q} + f$$
without the ESEEM term as the data was too noisy at higher temperatures to see the ESEEM. We discovered that $T_2$ in this material is largely invariant across the temperature range apart from a decrease from 6.35 to 4.71 μs between T = 5 K and T = 15 K. $T_2$ remains at approximately 4 μs until T = 60 K before it decreases to approximately 1 μs at T = 90 K and remains roughly constant until T = 150 K, Fig 5(d). The temperature invariance of $T_2$ across the temperature range could be a result of the relative lack of nuclear spins and methyl groups within the lattice which can easily promote decoherence in electronic spin-based systems [38]. The low value of $T_2$ across the temperature range could, however, stem from the non-zero natural abundance (14.31%) of $^{183}$W. Phase memory times are extremely sensitive to even incremental changes in nuclear spin concentration, as has been observed in studies examining the isotopic enrichment of diamond hosts for nitrogen vacancy centers [40].

## IV. CONCLUSIONS

We have grown single crystals of Ba$_2$CaWO$_{6-\delta}$ for the first time. Continuous wave and pulse EPR measurements confirm the presence of oxygen vacancies that create W$^{5+}$ defect centers in the system. Coherence studies indicate that these defects in Ba$_2$CaWO$_{6-\delta}$ are viable quantum bit candidates. Without significant optimization, the longitudinal relaxation time, $T_1$ in this material rivals that of top quantum bit candidates, and the transverse relaxation time, $T_2$ shows relative insensitivity to temperature across the measurement range. Remarkably, $T_1$ = 310 ms at T = 5 K, only decreasing upon population of low-lying phonon modes at 15 K and with the onset of local vibrational modes above 60 K. Work towards reducing the influence of these modes is ongoing. Specifically, we plan to slow phonon-mediated relaxation by designing systems with lighter elements with less spin-orbit coupling and systems with less susceptibility to octahedral tilting. Further we calculate the spin-phonon coupling of the higher frequency Debye 2 mode to be ~ 3 times as strong as the Einstein or the Debye 1 modes. The fact that all coupling strengths are not equal is chemically intuitive; our results show how to gain access to such information in a general, broadly applicable way. It also suggests that simply pushing all vibrational spectral weight to high frequencies is not the only way to design a high $T_1$ lifetime – instead, reducing the spin-phonon coupling to only the modes present at low energy (e.g. by making such coupling symmetry-forbidden), is the key. In short, complex oxides are viable hosts for quantum bit centers and the chemical control of oxygen vacancies can be used to introduce spin centers in the lattice. We further find that systematic materials design principles can be used to create qubits with long longitudinal relaxation times.

## ACKNOWLEDGEMENTS

This work was funded by the Platform for the Accelerated Realization, Analysis, and Discovery of Interface Materials (PARADIM), a National Science Foundation Materials Innovation Platform (NSF DMR-1539918). TMM acknowledges support of The David and Lucile Packard Foundation. Use of the Advanced Photon Source at Argonne National Laboratory was supported by the U. S. Department of Energy, Office of Science, Office of Basic Energy Sciences, under Contract No. DE-AC02-06CH11357. EPR studies were performed at the University of Illinois School of Chemical Sciences EPR lab (Urbana, IL) and at the National Biomedical EPR Center at the Medical College of Wisconsin (Milwaukee,

WI)(NIH P41 EB001980). TJP gratefully acknowledges the support of an NSF Graduate Research Fellowship (DGE-1324585). WAP and MS are grateful to Dr. Saul Lapidus of 11-BM for useful discussions related to data collection strategies. MS would like to thank Dr. Maxime Siegler, Dr. Thao T. Tran, Christopher M. Pasco for helping with single crystal data collection, analysis and synchrotron XRPD analysis respectively. TJP is grateful to Dr. Mark Nilges (UIUC) and Dr. Candice Klug (MCW) for assistance with EPR spectroscopy experiments.

**REFERENCES**

1. M. A. Nielsen, I. Chuang, *American Journal of Physics* **70**, 345 (2002).
2. R. P. Feynman, *Int. J. Theor. Phys.* **21 (6–7)**, 467–488 (1982).
3. M.H. Devoret, R.J. Schoelkopf, *Science* **339 (6124)**, 1169-1174 (2013).
4. M. N. Leuenberger, D. Loss, *Nature* **410 (6830)**, 789 (2001).
5. J. J. Pla, K. Y. Tan, J. P. Dehollain, W. H. Lim, J. J. Morton, D. N. Jamieson, A. S. Dzurak, A. Morello, *Nature* **489 (7417)**, 541 (2012).
6. Y. Morita, S. Suzuki, K. Sato, T. Takui, *Nat. Chem.* **3 (3)**, 197 (2011).
7. J. M. Zadrozny, J. Niklas, O. G. Poluektov, D. E. Freedman, *ACS Cent. Sci.* **1 (9)**, 488–492 (2015).
8. M. J. Graham, J. M. Zadrozny, M. Shiddiq, J. S. Anderson, M. S. Fataftah, S. Hill, D. E. Freedman, *J. Am. Chem. Soc.* **136 (21)**, 7623–7626 (2014).
9. G. Aromí, D. Aguilà, P. Gamez, F. Luis, O. Roubeau, *Chem. Soc. Rev.* **41 (2)**, 537–546 (2012).
10. F. Troiani, M. Affronte, *Chem. Soc. Rev.* **40 (6)**, 3119–3129 (2011).
11. P. C. Stamp, A. J. Gaita-Ariño, *J. Mater. Chem.* **19 (12)**, 1718–1730 (2009).
12. M. W. Doherty, N. B. Manson, P. Delaney, F. Jelezko, J. Wrachtrup, L. C. L. Hollenberg, *Phys. Rep.* **528**, 1–45 (2013).
13. W. F. Koehl, B. B. Buckley, F. J. Heremans, G. Calusine, D. D. Awschalom, *Nature* **479**, 84–87 (2011).
14. M. Widmann, S. Y. Lee, T. Rendler, N. T. Son, H. Fedder, S. Paik, L.P. Yang, N. Zhao, S. Yang, I. Booker, A. Denisenko, M. Jamali, S. A. Momenzadeh, I. Gerhardt, T. Ohshima, A. Gali, E. Janzén, J. Wrachtrup, *Nat. Mater.* **14**, 164–168 (2014).
15. D. J. Christle, A. L. Falk, P. Andrich, P. V. Klimov, J.U. Hassan, N. T. Son, E. Janzén, T. Ohshima, D. D. Awschalom, *Nat. Mater*. **14**, 160–163 (2014).
16. S. Nellutla, G.W Morley, J. van Tol, M. Pati, N.S. Dalal, *Phys. Rev. B* **78 (5)**, 54426 (2008).
17. M. Shiddiq, D. Komijani, Y. Duan, A. Gaita-Ariño, E. Coronado, S. Hill, *Nature* **531**, 348 (2016).
18. K. Bader, D. Dengler, S. Lenz, B. Endeward, S.D. Jiang, P. Neugebauer, J. van Slageren, *Nat. Commun.* **5**, 5304 (2014).
19. J. J. Morton, A. M. Tyryshkin, A. Ardavan, K. Porfyrakis, S. Lyon, D. Andrew, G. J. Briggs, *Chem. Phys.* **124 (1)**, 14508 (2006).
20. S. Takahashi, R. Hanson, J. van Tol, M. S. Sherwin, D. D. Awschalom, *Phys. Rev. Lett.* **101 (4)**, 47601 (2008).
21. P. L. Stanwix, L. M. Pham, J. R. Maze, D. LeSage, T. K. Yeung, P. Cappellaro, P. R. Hemmer, A. Yacoby, M. D. Lukin, R. L. Walsworth, *Phys. Rev. B* **82 (20)**, 201201 (2010).
22. S. Bertaina, S. Gambarelli, A. Tkachuk, I. Kurkin, B. Malkin, A. Stepanov, B. Barbara, *Nat. Nanotechnol.* **2 (1)**, 39 (2007).
23. T. J. Pearson, D. W. Laorenza, M. D. Krzyaniak, M. R. Wasielewski, D. E. Freedman, *Dalton Trans.* **47 (34)**, 11744–11748 (2018).
24. L. Tesi, E. Lucaccini, I. Cimatti, M. Perfetti, M. Mannini, M. Atzori, E. Morra, M. Chiesa, A. Caneschi, L. Sorace, R. Sessoli, *Chem. Sci.* **7 (3)**, 2074–2083 (2016).
25. M. Atzori, S. Benci, E. Morra, L. Tesi, M. Chiesa, R. Torre, L. Sorace, R. Sessoli, *Inorg. Chem.* **57 (2)**, 731–740 (2017).
26. N. Bar-Gill, L.M. Pham, C. Belthangady, D. Le sage, P. Cappellaro, J.R. Maze, M.D. Lukin, A. Yacoby, R. Walsworth, *Nature Comm.* **3, 858,** 1-6 (2012).
27. G.de Lange, Z. H. Wang, D. Rist, V. V. Dobrovitski, R. Hanson, *Science* **330**, 60–63 (2010).
28. B. Naydenov, F. Dolde, L. T. Hall, C. Shin, H. Fedder, L. C. L. Hollenberg, F. Jelezko, J. Wrachtrup, *Phys. Rev. B* **83**, 081201-1-4, 2011.
29. C. A Ryan, J. S. Hodges, D. G. Cory, *Phys. Rev. Lett.* **105**, 200402-1-4, (2010).
30. L. Gordon, J.R. Weber, J.B. Varley, A., Janotti D.D. Awschalom, C.G.V. Walle, *MRS BULLETIN* **38**, 802-807 (Oct 2013).
31. B. E. Day, N. D. Bley, H. R. Jones, R. M. McCullough, H. W. Eng, S.H. Porter, P. M. Woodward, P. W. Barnes, *Solid State Chem.* **185**, 107-116 (2012).
32. S. Stoll, A. Schweiger, *J. Magn. Reson.* **178(1)**, 42-55 (2006).
33. R.H. Lamoreaux D.L. Hildenbrand, *J. Phys. Chem. Ref. Data* **16**, 3, 419-443 (1987).
34. A.M. Glazer, *Acta Cryst.* **B28**, 3384 (1972).
35. A.P. Ramirez, G.R. Kowach, *Phys. Rev. Lett.* **80**, 4903–490 (1998).
36. J. Meng, Q. Lin, T. Chen, X. Wei, J. Li, Z. Zhang, *Nanoscale* **10**, 2908-2915, (2018).
37. K.M. Salikhov, S.A. Dxuba, A.M. Raitsimring, *Journal of Magnetic Resonance*, **42**, 255–276 (1981).
38. L. Berliner, S.S. Eaton, G.R. Eaton, *Biological Magnetic Resonance: Distance Measurements in Biological Systems by EPR,* (Kluwer Academic/Plenum Publishers, New York, 2000), Vol. **19**, p. 68-69, 120-121, 123.
39. S.K. Hoffmann, S. Lejewski, *Journal of Magnetic Resonance.* **252,** 49–54(2015).
40. G. Balasubramanian, P. Neumann, D. Twitchen, M. Markham, R. Kolesov, N. Mizouchi, J. Isoya, J. Achard,

J. Beck, J. Tissler, V. Jacques, P.R. Hemmer, F. Jelezko, J. Wrachtrup, *Nature Mater.* **8**, 383–387 (2009).